\documentclass[journal, 9pt]{IEEEtran}

\ifCLASSINFOpdf
\else
\fi
\usepackage{cite}
\usepackage{graphicx}
\usepackage{float}
\usepackage[small]{caption}
\usepackage{subcaption}
\usepackage{color,soul}
\usepackage{array}
\usepackage{amsmath}
\usepackage{booktabs}
\usepackage{footnote}
\makesavenoteenv{tabular}
\makesavenoteenv{table}
\usepackage{color}
\usepackage{amssymb}
\usepackage[linesnumbered,ruled,vlined]{algorithm2e}

\usepackage{algpseudocode}

\usepackage{setspace}

\begin{document}
\title{Deep-PowerX: A Deep Learning-Based Framework for Low-Power Approximate Logic Synthesis}

\author{Ghasem Pasandi, Mackenzie Peterson, Moises Herrera, Shahin Nazarian and Massoud Pedram \\
  Department of Electrical and Computer Engineering\\ University of Southern California (USC), Los Angeles, CA 90089.\\\{pasandi, mackenzp, herrerab, shahin, pedram\}@usc.edu

}

\markboth{2020 ACM/IEEE International Symposium on Low Power Electronics and Design (ISLPED '20), Boston, MA, USA. DOI: 10.1145/3370748.3406555}
{Pasandi et al.: Deep-PowerX: A Deep Learning-Based Framework for Low-Power Approximate Logic Synthesis}

\maketitle

\begin{abstract} 
This paper aims at integrating three powerful techniques namely Deep Learning, Approximate Computing, and Low Power Design into a strategy to optimize logic at the synthesis level. 
We utilize advances in deep learning to guide an approximate logic synthesis engine to minimize the  dynamic power consumption of a given digital CMOS circuit, subject to a predetermined error rate at the primary outputs. Our framework, Deep-PowerX\footnote{https://github.com/mackenzp/als}, focuses on replacing or removing gates on a technology-mapped network and uses a Deep Neural Network (DNN) to predict error rates at primary outputs of the circuit when a specific part of the netlist is approximated. The primary goal of Deep-PowerX is to reduce the dynamic power whereas area reduction serves as a secondary objective. Using the said DNN, Deep-PowerX is able to reduce the exponential time complexity of standard approximate logic synthesis to linear time.
Experiments are done on numerous open source benchmark circuits. Results show significant reduction in power and area by up to $1.47\times$ and $1.43\times$ compared to exact solutions and by up to 22\% and 27\% compared to state-of-the-art approximate logic synthesis tools while having orders of magnitudes lower run-time.
\end{abstract}

\IEEEpeerreviewmaketitle

\section{Introduction}
\label{Intro:sec}
Due to an ever increasing usage of portable devices such as cell phones, notebook computers, and personal digital assistants (PDAs), low-power and energy-efficient design of digital circuits and systems has gained a lot of attention. This is because of the growing need to increase the battery-based operation time of these devices by reducing their average energy consumption. Emerging applications in computer science and vision such as video and image processing \cite{huang2015surface}, fast search engines \cite{ranjan2016approximate,yang2003approximate}, deep learning and machine learning \cite{lecun2015deep, krizhevsky2012imagenet, pasandi2018TruthNet} have opened new opportunities for low-power and energy-efficient circuit and system design. These applications typically require a large amount of computation implying high power consumption. Fortunately, these computations can tolerate some degree of inaccuracy in their final results. An approximate computing paradigm enables us to take advantage of the power-accuracy trade-off.

Approximate computing is a computing technique, which although does not guarantee exactness, produces results with a sufficient level of accuracy that meets the application needs. This is done by relaxing the exact equivalency requirements between provided specifications and generated implementation results. 
The idea of approximate computing can be implemented at different levels of the design hierarchy. This paper presents a realization of the approximate computing approach by focusing on the logic synthesis step in the design flow. Logic synthesis has two phases (technology-independent and technology mapping) and is defined as the process of optimizing a given Boolean network and  mapping it to a gate level netlist while optimizing power, area, delay, or any other desired metric. 
There are multiple previous works detailing technology-independent optimization during logic synthesis, focusing on approximate solutions that maintain a constraint on either the error rate or error magnitude. Such works attempt to reduce the total area and/or critical path delay of the final approximate circuit.
These techniques mostly have a greedy nature and have a tendency to incur longer run-times  as well as high costs of ensuring that a satisfactory accuracy level is achieved after the approximation. Additionally, these techniques lack a learning process that would allow the framework to learn from the previous experiences in order to become more effective and run-time efficient.

In this paper, we present Deep-PowerX, a novel approach for Approximate Logic Synthesis (ALS) which utilizes machine learning algorithms to target minimization of the circuit power consumption. Deep-PowerX benefits from advances in deep learning by utilizing a Deep Neural Network (DNN) for fast calculation of error rate for an arbitrary netlist during the approximation process. During the training phase of Deep-PowerX, training data is generated and used for training the embedded DNN. Each training data vector includes features of a node which is to be approximated, with a projected maximum error rate at the primary outputs of the circuit. In the inference phase, Deep-PowerX receives as input, a mapped circuit and traverses the circuit in order to recommend replacements for each node. At each approximation step, Deep-PowerX consults with the embedded DNN, which is trained for predicting the error rate at outputs of the given netlist. If the predicted error rate is more than the predetermined error rate given by the user, that specific gate replacement for the node under consideration is abandoned; otherwise, it will be accepted.  

The embedded DNN in Deep-PowerX receives features of a target gate in a circuit and its surrounding gates as inputs and predicts the resultant maximum error rate at the primary outputs of the circuit. The predicted error rate is defined by the normalized Hamming distance between the exact and approximate truth tables of the primary outputs. Boolean difference calculus is applied to calculate the error rate at the primary outputs due to local gate approximations and is further used for DNN training and calibration during implementation. Experimental results demonstrate that Deep-PowerX achieves significant improvements in terms of run-time, power, and area savings over state-of-the-art ALS frameworks. Due to the ease of integration with industry standard synthesis tools, the  approximation is done after the technology mapping phase. We believe that this is the first paper to address the problem of approximate logic synthesis incorporating deep learning in the process of approximation.
\section{Background}
\label{back:sec}
\subsection{Probabilistic Error Propagation}
\label{PEP_sub_sec}
\begin{figure}[t]
\centering
\includegraphics[width=0.22\textwidth]{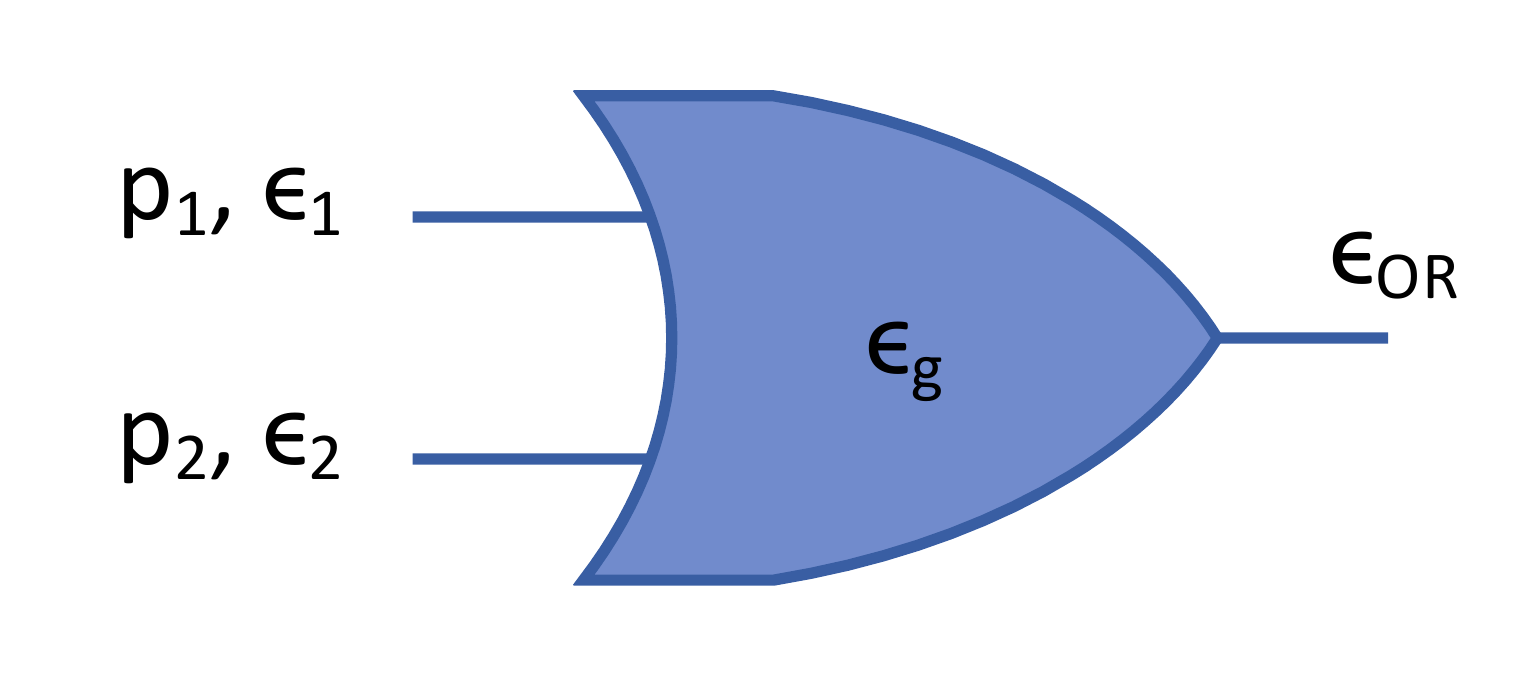}
\caption{Faulty 2-input OR gate with erroneous inputs. $\epsilon_1$ and $\epsilon_2$ are error probabilities at inputs, $p_1$ and $p_2$ are signal probabilities at inputs of this gate, and $\epsilon_g$ is intrinsic error probability of the gate. The error probability at the output, $\epsilon_{OR}$ is given by Eq. \ref{epsilon_OR2}.}
\label{OR2_fig}
\end{figure}

In \cite{mohyuddin2011probabilistic}, a probabilistic error propagation method using Boolean difference calculus is presented to calculate the error rate at the output of a logic gate. In this method, having as input the Boolean function of the gate, error probabilities at its inputs, and the intrinsic error probability of the gate itself, the probability of error at the output of this gate is calculated. For example, as shown in Fig. \ref{OR2_fig}, having intrinsic error probability of a 2-input OR gate, $\epsilon_g$, the signal probabilities at its inputs (probabilities for input signal to be 1), $p_1$ and $p_2$, while the input error probabilities are $\epsilon_1$ and $\epsilon_2$, the probability of error at the output of this gate is:
\begin{equation}
\label{epsilon_OR2}
    \epsilon_{OR} = \epsilon_g + (1-2\epsilon_g)(\epsilon_1(1-p_2)+\epsilon_2(1-p_1)-2\epsilon_1\epsilon_2(2p_1p_2-1))
\end{equation}
To calculate error rate at an output of a network resulting from error injection to one of its internal nodes, the above calculation should be done iteratively or recursively starting from this node and ending at the target output. 
\subsection{Design Space Complexity of Approximate Logic Synthesis}
\label{design_complx_sec}
The standard way of performing approximate logic synthesis by gate replacement involves going through all nodes and performing approximation on each and calculating the error rate at primary outputs resulting from this approximation. Therefore, the complexity analysis has two parts: node replacement, and error propagation. Let's assume that the given circuit is modeled by a Directed Acyclic Graph \emph{G=(V,E)}, where \emph{V} is the vertex set and \emph{E} is the edge set.
\begin{itemize}
    \item \textit{node replacement}: if there are up to \emph{k} possible replacements for a node \emph{n} in \emph{V}, the total number of possible approximations for the given circuit will have an upper bound of $k^n$.
    
    \item \textit{error propagation}: an approximated node injects error at its output which needs to be propagated throughout the circuit to find the error at primary outputs. Using the the error propagation method explained in Section \ref{PEP_sub_sec}, a Breadth-First-Search (BFS) with complexity of $O(m+n)$ is needed, where \emph{m} and \emph{n} are edge count ($|E|$) and node count ($|V|$), respectively. 
\end{itemize}
To verify that the error at primary outputs will be bounded by the given error constraint, error propagation should be done at each gate replacement iteration. Therefore, the total complexity will be: $O((m+n)\times k^n)$.
\subsection{Deep Neural Networks}
\label{DNN} 
A DNN has one input layer, one or more hidden layers and one output layer. Each layer is comprised of a group of neurons. Inputs of neurons in hidden layers travel through a non-linear activation function to learn any possibility of a complex relation that may be present between the input features and output classes. There are three main operations: The \emph{feedforward} operation computes activations and their derivatives  by using the weights, biases, and an activation function. The \emph{backpropagation} operation computes error values, while the \emph{update} operation modifies trainable parameters using a learning rate hyper-parameter. 
\section{Related Work} %
\label{prior:sec}
This section has the following logical flow: first we review a few papers on the topic of ALS, then we bring some others which are focused on optimizing power and energy during approximation. Finally, we will mention three papers that use machine leaning and deep learning in logic synthesis. Our paper has a flavor of all because it uses deep learning in ALS and targets power minimization. 

Wu and Qian \cite{wu2016efficient} used approximation of factored forms of Boolean expressions for each node to come up with efficient approximation for the whole circuit. They have implemented two versions namely single-selection and multi selection with better QoR for the former and lower run-time for the latter one. 
Hashemi \textit{et al.} \cite{blasys1} proposed a Boolean Matrix Factorization (BMF) method to provide approximation on the Boolean-level representation of a given circuit. Additionally, a decomposition method of subcircuits was proposed to provide a trade-off between the required accuracy and the circuit complexity.
Zhou \textit{et al.} in \cite{dals} proposed a delay-driven ALS framework that utilizes an And-Inverter Graph (AIG) representation of a given circuit in order to optimize the circuit's performance.

For power and energy minimization in approximate logic synthesis, there are several papers in the literature. 
Swagath \textit{et al.} \cite{venkataramani2013substitute} presented a framework for approximate logic synthesis targeting area and power minimization by functional approximations inside the circuit through logic gate removal and function simplification in an iterative way until the error constraint is reached. 
Schlachter \textit{et al.} \cite{gate_level_prunning} proposed a gate-level pruning method to perform approximation on arithmetic circuits found in 
functional blocks of discrete cosine transformation units that is used in image and video processing. The authors obtained a trade-off between accuracy and power consumption using this approach.
In \cite{VADER}, Zervakis \textit{et al.} introduced a  voltage-driven functional approximation method to perform gate pruning after synthesis.
Their experimental results were applied to adders and multipliers, providing improvements on both energy and area.

Regarding usage of machine learning and deep learning algorithms in logic synthesis, there are some papers published recently; 
Q-ALS, a reinforcement learning-based framework for approximate logic synthesis was presented in \cite{Q-ALS}. Q-ALS learns the maximum error rate tolerable by each node in order to optimize the circuit for delay and then area while adhering to a predetermined error rate at primary outputs. Q-ALS is the first framework that formulates the technology mapping problem as a reinforcement-learning problem and gives solid definitions for state, action, and reward functions.
Yu \textit{et al.} \cite{yu2018developing} proposed an exact synthesis flow utilizing a Convolutional Neural Network (CNN) targeting elimination of human experts from the whole process. The authors could generate the best designs for three large scale circuits, beating the state-of-the-art logic synthesis tools. In \cite{hosny2019drills}, a deep reinforcement learning approach for exact logic synthesis is presented. The authors have used A2C reinforcement learning algorithm to determine the order of applying optimization commands (among a few candidate commands) to a given circuit for achieving better QoR. Similar to \cite{yu2018developing}, in \cite{hosny2019drills}, the goal is to remove the human guidance and expertise from the process of logic synthesis. In this paper, we present Deep-PowerX, which provides low-power and area-efficient approximate logic solutions while benefiting from state-of-the-art deep learning algorithms to offer significant improvements on QoR (power, area, delay, and run-time). 
\begin{figure*}[t]
\centering
\includegraphics[width=0.75\textwidth]{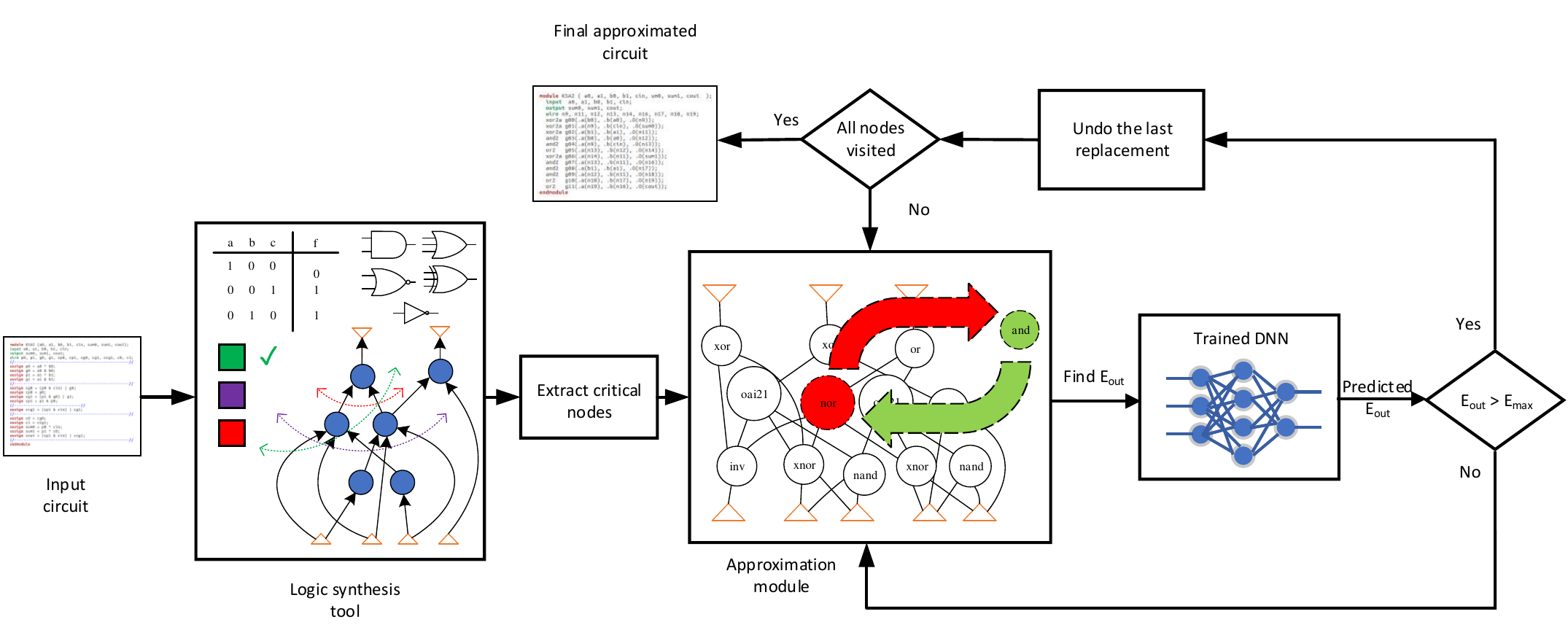}
\caption{Deep-PowerX Framework. First, the input circuit is mapped using the cut-based technology mapping module of ABC \cite{synthesis2011abc}. Next, depending on the optimization mode (e.g. power), critical nodes are extracted and passed to the approximation module; this module replaces a node and consults with the trained DNN to predict the worst error rate at primary outputs. If this worst error rate is not violating the predetermined error constraint by a user, the approximation module accepts the replacement, otherwise, it will undo the last replacement and moves to another node. The approximation process ends when either the error budget is fully consumed or all nodes are visited. }
\label{overallapp}
\end{figure*}        
\section{Proposed Framework: Deep-PowerX}
\label{prop:sec}
Our proposed framework for minimum-power approximate logic synthesis is called \emph{Deep-PowerX}. The first word refers to deep learning which is employed for error calculation and for providing recommendations for gate replacements; \emph{Power} denotes the main optimization objective, whereas \emph{X} refers to the approximate computing paradigm of our framework. 
Fig. \ref{overallapp} illustrates the Deep-PowerX flowchart. In this implementation of Deep-PowerX, power/area minimization algorithms and a DNN are utilized to produce the best approximate netlist solution for a given Boolean network. The embedded DNN in Deep-PowerX is trained using some sample approximated networks having the error rates provided at their primary outputs. These error rates are in turn calculated by applying the Boolean difference calculus to calculate the normalized Hamming distance between the truth table of the exact network and its approximate implementation (cf. Section \ref{training_data_sub_sec}). The replacement algorithm first attempts to replace \emph{critical power nodes}\footnote{Critical power nodes are those with the highest switching powers.} and a portion of their immediate fanout nodes with simpler nodes and then goes on to the area minimization phase, while continuously consulting the embedded DNN to ensure that the error rate at primary outputs is kept within the specified user constraints. Deep-PowerX is trained to minimize the total switching power consumption during the approximation process.
Deep-PowerX has three main phases: \emph{training}, \emph{inference}, and \emph{testing}.
\begin{algorithm} [t]
{\small
\caption{Deep-PowerX Training Data Generation}\label{algo:training}
\DontPrintSemicolon 
\KwIn{Set of training networks, technology library } 
\KwOut{X: training data vectors, Y: labels}
X = \{\}; Y = \{\};  \\
\For{each training network $N$}{
    $N_{MAP}$ = Map $N$ with the given technology library.\\
    \For{each node $n$ in $N_{MAP}$}{
        $X_1$ = Extract features of $n$. \\ 
        temp = $X_1$; \\
        \For{each possible replacements for $n$}{  
            $\epsilon$ = Calculate local error resulting from replacement.\\
            $X_1$ = temp; $X_1$.append($\epsilon$); \\
            $X_1$.append(errors at immediate fanins of $n$) \\
            $E_{out}$ = Calculate output error rate recursively. \\
            X.append($X_1$); Y.append($E_{out}$);
        }
    }
}
\Return{X, Y}
}
\end{algorithm}
\subsection{Training}
\label{training_data_sub_sec}
The training phase is comprised of two steps: \emph{training data generation}, and \emph{performing training of the on-board DNN using this data}.\\ 
First, in an offline step, Deep-PowerX generates the training data to be used for training its embedded DNN. For a training network, $N=(V,E)$, Deep-PowerX traverses over its nodes and calculates the error that can be injected into the network by replacing some node with a gate from the library. This error is calculated by finding the Hamming distance between truth tables of this node and that of the replacing gate. Next, using a probabilistic error propagation method as described in Section \ref{PEP_sub_sec}, the corresponding error rates at primary outputs for this gate replacement are estimated. Finally, some relevant features of the node under analysis including the node type, replacing gate type, local error due to replacement, number of fanouts, number of fanins, logic level, logic depth of the circuit, and all fanin and fanout node features (up to a certain depth limit) are extracted. These features comprise one training data point and the estimated error rate will be its label. This process is continued for all nodes in all training networks. When the training data is generated, it will be used for training the embedded DNN in the next step. 
The training data generation is shown in Algorithm \ref{algo:training}. Inputs of the algorithm are training networks and a technology library, and its outputs are a list of training vectors together with their corresponding error rates as labels. Note that local errors resulting from a node replacement is also included in the feature vectors. Our experiments show that this helps the DNN converge much faster and provides more accurate predictions.

In the second step, the generated training data is used for training the on-board DNN. Dimensions of the input layer are determined by a node with the highest number of fanins/fanouts in a dense training network. Note that since we include parts of features of nodes within the fanin/fanout cone of node $n$ in its feature vector, the length of the feature vector for node $n$ will increase if it has many nodes in its fanin/fanout cones. Based on our experiments on our training networks, we set 93 as the maximum value for length of a feature vector. For nodes that have a shorter feature vector size, we append zeros to their feature vectors to bring them into this standard size. The DNN in Deep-PowerX is comprised of two fully connected hidden layers with 400 and 300 neurons in the first and second layers, respectively. Also, this DNN has 51 neurons in its output layer. The number of neurons on the output layer is chosen to achieve the desired accuracy on the error prediction. 
We have used \emph{Adam} optimizer, \emph{binary cross entropy} as the loss function, and 30 as the number of \emph{epochs}.
The DNN model and training parameters are shown in Table \ref{model_params}.
\begin{table}[t]
\caption{The DNN model and parameters used in Deep-PowerX.} 
\begin{center}
\begin{tabular}{|c|c|}
\hline
{Network structure} & {93-400-300-51}\\
\hline
{Learning rate} & {0.001} \\
\hline
{Optimizer} & {Adam}\\
\hline
{Loss function} & {Binary cross entropy}\\
\hline
{Epochs} & {30}\\
\hline
\end{tabular}
\label{model_params}
\end{center}
\end{table}
\subsection{Inference}
\label{infer_sub_sec}
\begin{algorithm} [t]
{\small
\caption{Deep-PowerX Inference}\label{algo:per_approx}
\DontPrintSemicolon 
\KwIn{Input circuit $G=(V,E)$,  Technology library,  Max. error constraint ($E_{max}$)}
\KwOut{Approx. circuit $G_X=(V_X,E_X)$}
\tcp{\small{Initialization:}}
$G_X$ = Copy $G$.\\ 
$V'$  = Extract top 20\% of active nodes of $V_X$.\\ 
$V''$  = Extract top 20\% immediate fanouts of nodes in $V'$.\\ 
set $E_{out}$ to 0.0, and copy technology library to \emph{Lib}. \\
\tcp{\small{Minimizing power:}}
Lib1 (Lib2) = Sort Lib in ascending order by gates' out (in) cap. \\
opt\_mode = power; \\
$V'''$ = $V' \cup V''$ \\
\For{each node $n$ in $V'''$}{
    \If{opt\_mode is power}{
        \If{$n \in V'$}{
            Lib = Lib1;
        }
        \ElseIf{$n \in V''$}{
            Lib = Lib2;
        }
    }
    \For{each gate $g$ in Lib}{
        Replace $n$ with $g$; Utilize DNN to infer $E_{pred}$\\
        \If {replacement is accepted}{
            Update $E_{out}$ with $E_{pred}$. \\
            Break;
        }
        \Else{
            Undo the last gate replacement. 
        }
    }
    \If{$E_{out} > E_{max}$}{
        Undo the last gate replacement.\\ 
        Break;
    }
}
\tcp{\small{Minimizing area:}}
\If{$E_{out} < E_{max}$}{
    Lib = Sort Lib in ascending order by gates' area.\\
    opt\_mode = area;\\
    \tcp{\small{Copy nodes of approx. circuit in power minimization phase:}}
    $V'''$ = $V_X$; Go to line 8.
}
\Return{$G_X$}
}
\end{algorithm}
First, the switching activities of nodes are extracted using a probabilistic simulation-based method \cite{jang2009power,iman1998logic}. Then these nodes are sorted in descending order based on their switching activities and the top 20\% of critical power nodes are extracted. Next, by traversing the circuit level-by-level starting from primary inputs, critical power nodes and also 20\% of their immediate fanout nodes (that have not been previously replaced) are considered as candidates for approximation. Critical power nodes are replaced with gates that have smaller output capacitance and their immediate fanout nodes are replaced with gates with smaller input capacitance. This way the effective switching capacitance of critical nodes will be reduced, resulting in reduction in total switching power of the circuit. A user can select to preserve the best critical delay of the circuit. In this case, the gate replacement will be done only for nodes which are not on critical delay path and/or they will be replaced only if this replacement does not results in increasing the critical path delay.

A candidate node's features are extracted and are given to the DNN to predict the error rate at the primary outputs as a consequence of approximating the node. This process will continue until the critical power nodes (and 20\% of their immediate fanouts) are all have either been completely replaced or the error rate constraint has been violated. Next, if there is still room for additional approximations within the network, power optimization will continue but with gate removal instead of replacement. The power optimization algorithm will continue for as long as the error rate is within the user provided error rate constraint. If this error rate is violated, the last replacement will be undone and the area optimization algorithm will start. The area optimization algorithm works by starting at the first level of the network and by replacing each gate in that level with the lowest cost available gate in the library in terms of area. This will work level by level until finally the error rate constraint is reached or the entire circuit has been traversed. Algorithm \ref{algo:per_approx} shows the pseudo code for the inference phase of Deep-PowerX. Lines 1-5 are for initialization, lines 6-23 are for power minimization, and lines 24-27 are for area minimization.

After the network has been approximated, it will be exported and mapped again using ABC \cite{synthesis2011abc}. This allows the network to be further optimized. It is important to note that if the user wanted to prioritize area over power savings, the execution order of the algorithms could be swapped. This would dedicate a bigger portion of the error budget to the area optimization algorithm.

\subsection{Testing}
\label{Test_subSec}
We have used 60\% of the generated data as in Section \ref{training_data_sub_sec} for training, 20\% for validation, and 20\% for testing. We obtained a high accuracy of 98\% on the test data. This means that for 98\% of the cases, the embedded DNN in Deep-PowerX could predict the correct value for error rate at primary outputs of an unseen circuit, given a random approximation on any internal nodes of this circuit. This therefore confirms that our framework provides good generalization of learned knowledge from training networks to apply to unseen test networks. 
\subsection{Design Space Complexity}
\label{action_sub_sec}
In the inference phase, Deep-PowerX consults with the embedded DNN in order to find the maximum error rate at primary outputs resulting from an approximation. This is done once per each approximation iteration. Therefore, the $O(m+n)$ complexity of error propagation as in the standard way of performing approximation (Section \ref{design_complx_sec}) is reduced to a constant time. Also, Deep-PowerX replaces a candidate gate with another one from the library (which has a fixed gate count) with smaller output (input) capacitance. Finding such a gate and performing a replacement are done in a constant time. Therefore, the complexity of node replacement in Deep-PowerX is $O(k'\times n)$, where $k'$ is a constant. Given the constant time complexity for error estimation in Deep-PowerX, the total complexity will be $O(k''\times n)$, where $k''$ is another constant. The complexity of area minimization phase is the same, hence, the total complexity of Deep-PowerX is $O(n)$. 
\section{Experimental Results}
\label{exp:sec}
\begin{figure}[t]
\centering
\includegraphics[width=0.45\textwidth]{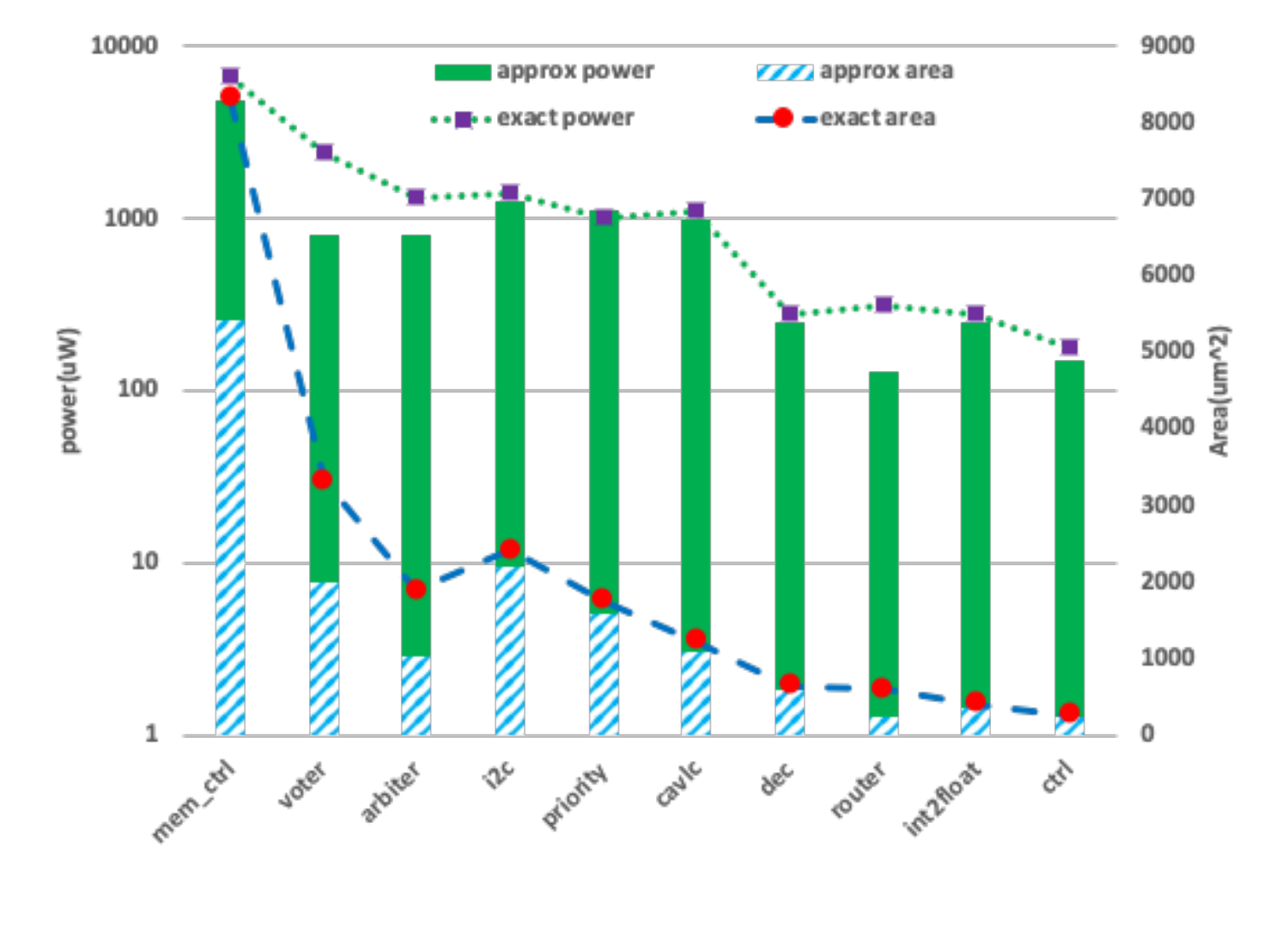}
\caption{Power ($\mu W$) and area ($\mu m^2$) results (both exact and approximate) for EPFL random\_control benchmark circuits. For better exhibition purposes, data for the three left most benchmarks is scaled down by a factor of 10. The left Y-axis is for power and the right Y-axis is for area. On average, Deep-PoweX reduced power and area by 49\%, and 41\%, respectively. }
\label{EPFL_power_fig}
\end{figure}        
\begin{table}[t]
\centering
\footnotesize\setlength{\tabcolsep}{2.5pt}
\scriptsize
\caption{Comparing amounts of reduction in power consumption and total area of a few benchmark circuits approximated by an state-of-the-art approximate logic synthesis tool, SASIMI \cite{venkataramani2013substitute}, and also by our Deep-PowerX framework. }
\begin{tabular}{@{}ccccccc@{}}
\toprule
\textbf{Circuit}  & Area ($\mu m^2$)  & Power ($\mu W$)  &  \multicolumn{2}{c}{SASIMI\cite{venkataramani2013substitute}}    & \multicolumn{2}{c}{Deep-PowerX}\\ 
 & & &area (\%)  & power (\%) & area (\%) & power (\%) \\
\midrule
KSA  & 1429.81	&910.2	&16.3	&14.79	&27.5&	26.4 \\
\midrule
c880    &    639.3	&335.84	&13.1	&18.03	&22.4	&28.4 \\
 \midrule
c1908       & 858.51	&583.3	&13.8	&22.9	&21.7	&45.38 \\ 
\midrule
c2670      & 1355	&851.54	&5.09	&15.68	&32.6	&29.6 \\ 
\midrule
c3540     &  1934.74	&1212.04	&21.94	&19.72	&17.6	&17.5\\ 
\midrule
c7552      &  3970.43	&3168.52	&12.79	&19.18	&13.5	&21.5 \\ 
\midrule
AVG  & 1697.96	&1176.90	&13.83	&18.38	&22.55	&28.13  \\
\bottomrule
\label{mcnc_comp_table}
\end{tabular}
\end{table}
\begin{figure}[h]
\centering
\includegraphics[width=0.4\textwidth]{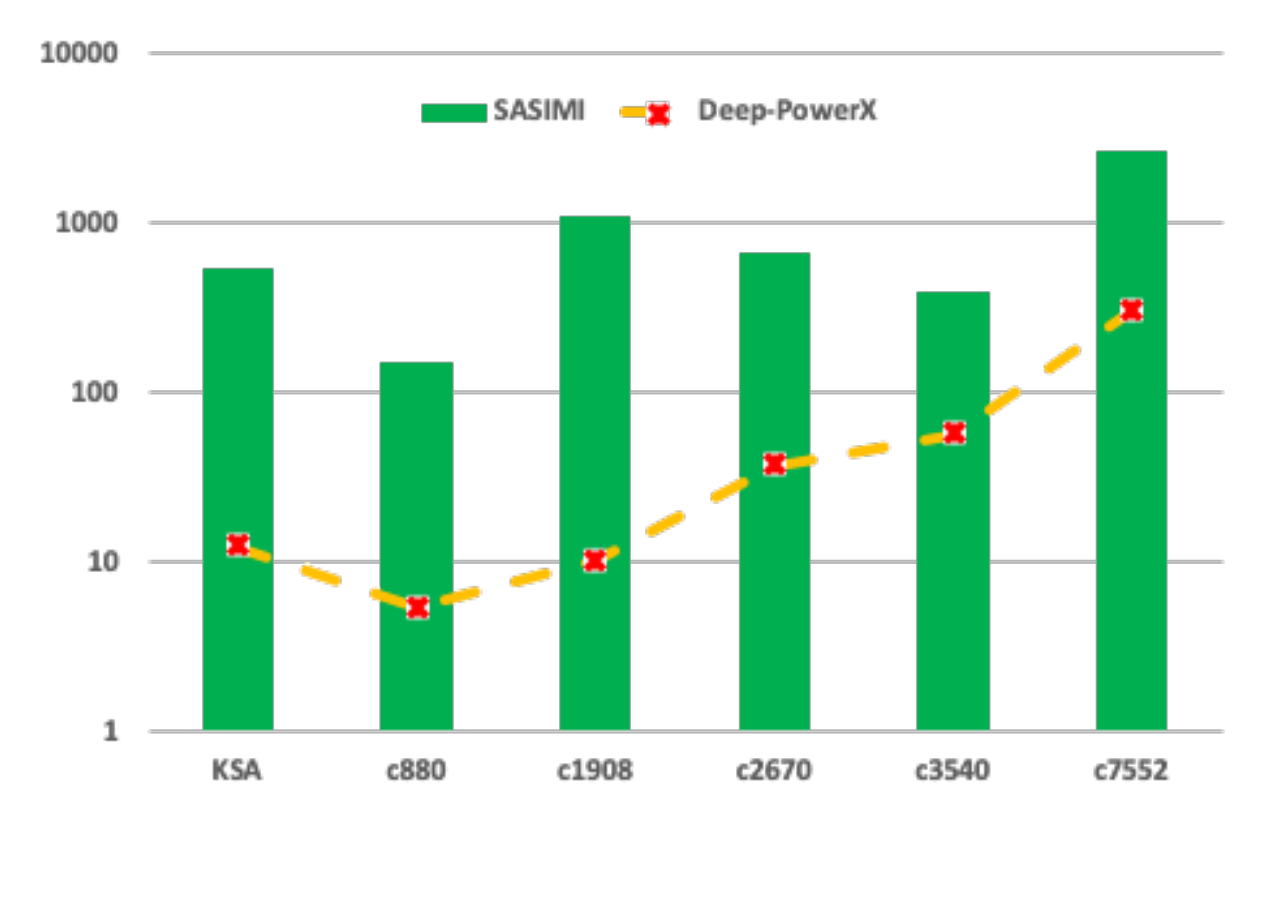}
\caption{Comparing run-time (s) for a few benchmark circuits (the same circuits as in Table \ref{mcnc_comp_table}) approximated by an state-of-the-art approximate logic synthesis tool, SASIMI \cite{venkataramani2013substitute}, and also by our Deep-PowerX framework (when it is in power+area optimization mode). }
\label{mcnc_runtime_fig}
\end{figure}        
\begin{figure}[t]
\centering
\includegraphics[width=0.5\textwidth]{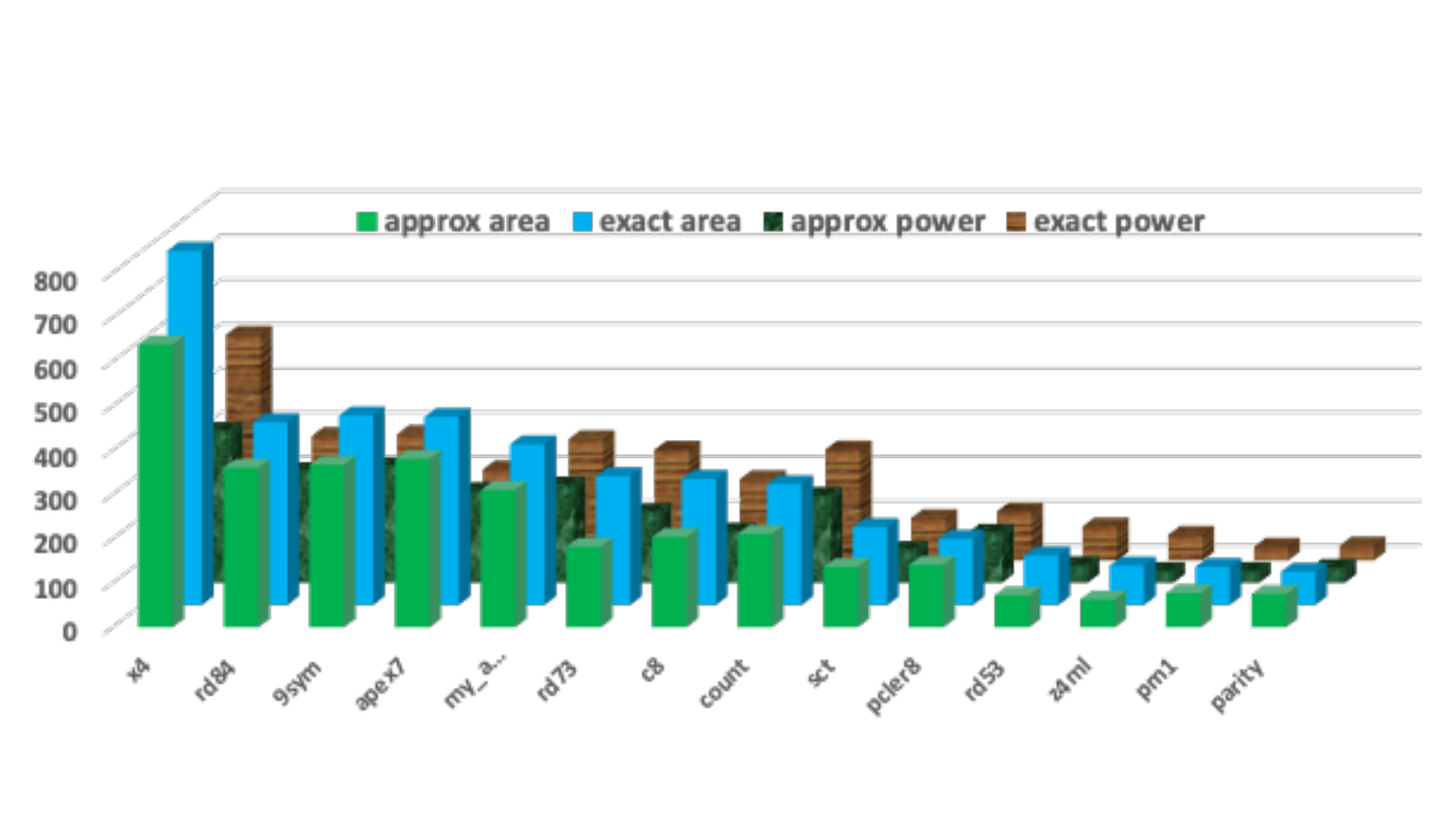}
\caption{Power ($\mu W$) and area ($\mu m^2$) results (both exact and approximate) for MCNC benchmarks. On average, Deep-PoweX reduced power and area by 37\%, and 27\%, respectively.}
\label{MCNC_power_fig}
\vspace{-1em}
\end{figure}        
We implemented our framework such that a user can select to perform power+area or delay+area optimization. In the latter case, instead of performing approximation on critical power nodes, those nodes on the critical delay path of the circuit will be replaced with faster gates from the technology library. At the end of this section, some experimental results for the delay+area mode will be presented. 
Similarly to previous ALS frameworks, the error rate constraint at primary outputs of the circuit was set to 5\%. This helps us better evaluate our framework. A 45nm ptm \cite{PTM} library was used to generate technology mapped netlists from circuits contained in benchmark suites such as ISCAS85 \cite{iscas} and MCNC \cite{yang1991logic}, and EPFL \cite{EPFL_bench}.
A subset of these combinational circuits were used in previous ALS frameworks and have been selected for direct comparison. First, we experimented on EPFL random\_control benchmark suite. This benchmark suite contains very large circuits such as \emph{mem\_ctrl} with 1204/1231 I/Os, 47110 nodes, and 93945 edges which are good candidates for evaluating the scalability of our framework. Fig. \ref{EPFL_power_fig} shows both exact and approximate power and area results for these circuits. Deep-PowerX reduced the power consumption by up to $1.47 \times$ and area by up to $1.43 \times$ for the \emph{router} circuit with a maximum error rate of 4.17\% at its primary outputs as compared to the exact circuit solution. Also, we saw a 36\% and 53\% reduction in power and area for the \emph{mem\_ctrl} circuit which demonstrates excellent scalability of Deep-PowerX for large circuits. On average for the 10 benchmark circuits in EPFL random\_control suite, Deep-PowerX reduced the power consumption and the area by 49\% and 41\%, respectively.
\begin{table}[t]
\centering
\footnotesize\setlength{\tabcolsep}{2.5pt}
\scriptsize
\caption{Comparing different state-of-the-art ALS tools in term of total area reduction when experimenting on a few ISCAS85 benchmark circuits.}
\label{my-label}
\begin{tabular}{@{}ccccc@{}}
\toprule
\textbf{Circuit}           & \textbf{SASIMI \cite{venkataramani2013substitute}}   & \textbf{Selection-based \cite{wu2016efficient}}  & \textbf{Q-ALS\cite{Q-ALS}} & \textbf{Deep-PowerX}  \\ 
 & (\%)  & (\%) & (\%) & (\%) \\
\midrule
c880    &    11.4     &  11.7 & 13.6 & 59.6 \\
 \midrule
c1908       & 39.0    &   40.2  & 39.5 &  3.6  \\ 
\midrule
c2670      & 28.6  &   32.7 & 33.3  & 96.8 \\ 
\midrule
c3540     &  2.5     &  3.5 &    4.5   &  3.4 \\ 
\midrule
c5315     &  1.9   &  1.9 &  37.9   &   67.2 \\ 
\midrule
c7552      &  5.2    &  5.9  & 38.0  &  2.4 \\ 
\midrule
AVG & 14.76  &15.98   &27.8    &28.9 \\
\bottomrule
\label{delay_area_comp_table}
\end{tabular}
\end{table}
\begin{figure}[t]
\centering
\includegraphics[width=0.4\textwidth]{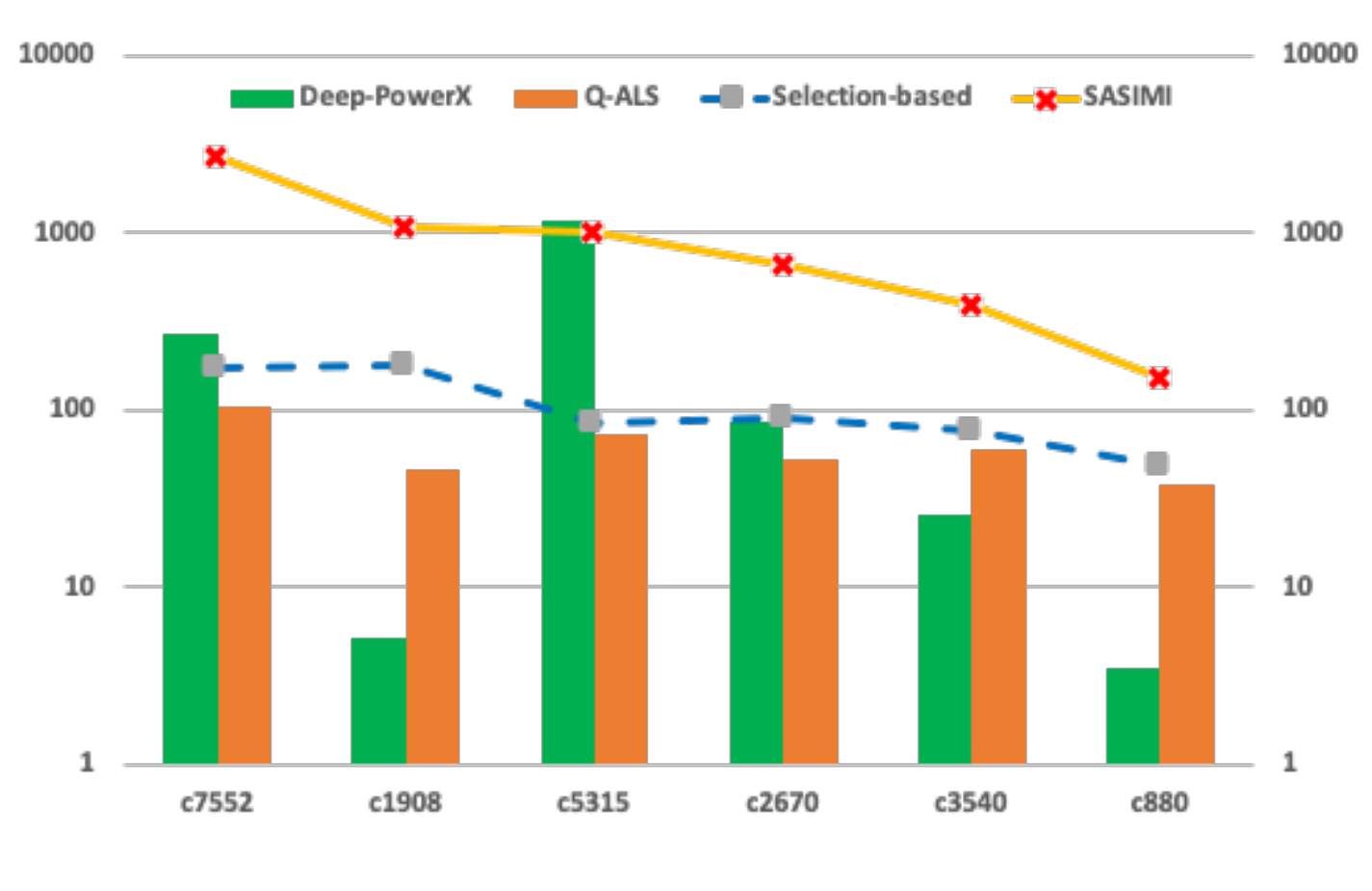}
\caption{Run-time comparison among ALS frameworks. The unit for time is second. It can be clearly seen that Deep-PowerX improves run-time significantly on average compared to SASIMI and selection-based frameworks. However, when it is compared with Q-ALS, the run-time is improved only on half of circuits.}
\label{run_time_fig}
\end{figure}        

We also experimented on MCNC benchmark suite. As in Fig. \ref{MCNC_power_fig}, Deep-PowerX could provide average savings of 37\% and 27\% on power and area compared with the exact solutions having the same error rate of 5\% at primary outputs.  
We also compared Deep-PowerX with one of the most powerful ALS frameworks (SASIMI \cite{venkataramani2013substitute}) that also offers optimizations for power and area. Therefore, this can be a good measure to evaluate our framework and see where it stands with respect to the state-of-the-art tools. In the SASIMI paper, power and area savings are reported for several open source circuits. We extracted power and area results for the same circuits using Deep-PowerX. Table \ref{mcnc_comp_table} lists these results for both SASIMI and Deep-PowerX. As seen in this table, only for \emph{c3540} circuit SASIMI performs better, but for the rest of the circuits, Deep-PowerX is better. On average for six circuits in Table \ref{mcnc_comp_table}, Deep-PowerX provides 22.55\% and 28.13\% savings on area and power respectively, while the amount of average savings for SASIMI is less with 13.83\% area savings and 18.38\% power savings. Fig. \ref{mcnc_runtime_fig} shows run-time for the same circuits listed in Table \ref{mcnc_comp_table}. Deep-PowerX experiences a significant amount of $34.42 \times$ average run-time savings when compared with SASIMI for the said benchmark circuits. 
For the delay+area optimization mode and again to compare with state-of-the-art ALS tools, we experimented on a few benchmark circuits with available results in three other ALS tools namely SASIMI, Selection-based \cite{wu2016efficient}, and Q-ALS\cite{Q-ALS}. Table \ref{delay_area_comp_table} shows the results. On average, Deep-PowerX could provide 14.14\% better area saving compared with SASIMI, and 12.92\% more reduction compared with selection-based approach in \cite{wu2016efficient}. Deep-PowerX could even improve the average area saving of Q-ALS, another powerful learning-based ALS tool, by a slight margin of 1.1\%. 
Fig. \ref{run_time_fig} shows the comparison of run-times of Deep-PowerX (when it is in delay+area optimization mode), Q-ALS, SASIMI and selection-based. As seen in this figure, Deep-PowerX improves run-time on all but the c5315 circuit when compared to SASIMI and selection-based frameworks. However, when it comes to Q-ALS, Deep-PowerX could improve the run-time for only half of the circuits. 
\section{Conclusion}
\label{conc:sec}
We presented Deep-PowerX, a DNN based approximate logic synthesis framework. Deep-PowerX has two optimization modes, namely power+area and delay+area. In the first optimization mode, nodes with highest switching activities and a portion of their immediate fanouts are approximated. In the second mode, nodes in the critical delay path are replaced with faster gates. In both modes, an embedded pre-trained DNN is used for guiding the synthesis tool to stick to a predetermined error rate at primary outputs. At the end of both modes, area minimization is performed in case additional approximation is possible. Experimental results on numerous circuits confirm significant improvements on QoR (power, area, delay, and run-time) for Deep-PowerX when compared to exact solutions and also state-of-the-art approximate logic synthesis tools.
\section*{Acknowledgement}
The research is supported in part by a grant from the Software and Hardware Foundations (SHF) program of the National Science Foundation. The authors would like to thank Souvik Kundu from the University of Southern California (USC) for his help in implementations used in this paper. 
\ifCLASSOPTIONcaptionsoff
  \newpage
\fi
\bibliographystyle{IEEEtran}
\bibliography{IEEEabrv,Deep-PowerX}
\end{document}